\documentclass[conference]{IEEEtran}
\IEEEoverridecommandlockouts
\usepackage{cite}
\usepackage{amsmath,amssymb,amsfonts}
\usepackage{algorithmic}
\usepackage{algorithm}
\usepackage{graphicx}
\usepackage[utf8]{inputenc}
\usepackage{siunitx}
\usepackage{booktabs, makecell}
\ifCLASSOPTIONcompsoc
    \usepackage[caption=false, font=normalsize, labelfont=sf, textfont=sf]{subfig}
\else
\usepackage[caption=false, font=footnotesize]{subfig}
\fi
\usepackage{nomencl}
\usepackage{multirow,tabularx}   
\usepackage{etoolbox}
\usepackage{textcomp}
\usepackage{xcolor}
\usepackage{verbatim}
\usepackage{mathtools}
\usepackage{mdwtab}
\usepackage{array}
\makeatletter
\newcommand*{\rom}[1]{\expandafter\@slowromancap\romannumeral #1@}
\makeatother

\begin{document}

\title{A Novel Data Segmentation Method \\for Data-driven Phase Identification\\}

\author{\IEEEauthorblockN{Han Pyo Lee, Mingzhi Zhang, \\ Mesut Baran, Ning Lu}
\IEEEauthorblockA{\textit{Dept. of Electrical and Computer Engineering}\\
\textit{North Carolina State University}\\
\textit{Raleigh, NC 27606, USA} \\
\{hlee39, mzhang33, baran, nlu2\}@ncsu.edu}
\and
\IEEEauthorblockN{PJ Rehm}
\IEEEauthorblockA{\textit{ElectriCities of North Carolina Inc.}\\
Raleigh, NC 27604, USA \\
prehm@electricities.org}
\and
\IEEEauthorblockN{Edmond Miller P.E., \\ Matthew Makdad P.E.}
\IEEEauthorblockA{\textit{New River Light and Power (NRLP)}\\
Boone, NC 28607, USA \\
\{millerec1, makdadmj\}@appstate.edu}
}

\maketitle
\bstctlcite{IEEEexample:BSTcontrol}
\begin{abstract}
This paper presents a smart meter phase identification algorithm for two cases: meter-phase-label-known and meter-phase-label-unknown. To improve the identification accuracy, a data segmentation method is proposed to exclude data segments that are collected when the voltage correlation between smart meters on the same phase are weakened. Then, using the selected data segments, a hierarchical clustering method is used to calculate the correlation distances and cluster the smart meters. If the phase labels are unknown, a Connected-Triple-based Similarity (CTS) method is adapted to further improve the phase identification accuracy of the ensemble clustering method. The methods are developed and tested on both synthetic and real feeder data sets. Simulation results show that the proposed phase identification algorithm outperforms the state-of-the-art methods in both accuracy and robustness.
\end{abstract}


\section{Introduction}
Maintaining accurate distribution topology information is crucial for distribution circuit analysis.  However, many topology information are registered into the customer information systems manually, making it susceptible to human errors. In addition, distribution circuits change frequently due to upgrades, reconfiguration, and equipment replacements. Therefore, it is also critical to keep the topology information, especially the phase of distribution transformers and customer meters, up-to-date.

However, the sheer number of distribution transformers and customers connected to the distribution system makes it impossible to manually check for data entry errors. Fortunately, metering technologies have advanced substantially over the years. In many areas, mechanical meters have been replaced by smart meters that record electricity use in 15- or 30- minute intervals. Many smart meters also provide voltage measurements. Thus, using smart meter data for phase identification becomes an attractive option.

In general, there are two approaches for phase identification: directly-measuring and data-driven. \textit{Directly measuring} approaches include signal-injection methods \cite{shen2013three} and micro-synchrophasor (PMU) methods \cite{wen2015phase}. In \cite{therrien2021assessment}, Therrien \emph{et al.} summarized the state-of-the-art of phase identification methods. \textit{Data-driven} approaches include real power-based methods \cite{arya2013inferring}, voltage-based methods \cite{pezeshki2012consumer}, and machine learning-based methods \cite{mitra2015voltage, wang2016phase, hosseini2020machine, foggo2019improving, blakely2020phase}. The main disadvantage of directly measuring is that it is more costly than using data-driven methods because of the additional equipment required and labor cost incurred. 

In recent publications applying the data-driven approach, machine learning-based methods are widely used. In \cite{mitra2015voltage}, Mitra \emph{et al.} applied k-means clustering using smart meter voltage data. In \cite{wang2016phase}, Wang \emph{et al.} proposed the advanced k-means clustering, which includes PCA and must-link constraints. In \cite{hosseini2020machine}, Hosseini \emph{et al.} proposed the use of a high-pass filter to first remove the low-frequency components in the time series load profiles before applying a modified k-means clustering algorithm. Supervised machine learning approach is proposed by Foggo \emph{et al.} in \cite{foggo2019improving}. This approach finds a constrained function of the voltage time series for correctly predicting the phase connections of a representative set. Then, the algorithm is applied to the remaining customers to obtain all of the phase connections on the feeder. The disadvantage of this approach is that it requires manual verification of phases and retraining is needed when applying to different feeders. In \cite{blakely2020phase}, Blakely and Reno proposed to use ensemble clustering to obtain final clusters by spectral clustering. A co-association matrix is generated using the similarity of ensemble clusters calculated by spectral clustering. In this method, data windows of time-series voltages were used as input. 

There are two main disadvantages of the state-of-the-art data-driven based methods: inefficient use of data and insufficient use of understandings derived from physics-based models. Thus, in this paper, we propose a novel data segmentation method. By applying circuit analysis, we prove the existence of voltage correlation deterioration phenomena between pairs of 1-phase customers on the same phase. This provides the theoretical foundation for the proposed data segmentation algorithm, where the low power and minimum duration threshold are used to extract only highly-correlated voltage data segments for computing the voltage correlation matrix required for hierarchical clustering. When the phase labels are unknown, we propose to use a Connected-Triple-based Similarity (CTS) method for further improving the phase identification accuracy of the ensemble clustering method. 

Our contributions are two-fold. First, for the first time in literature, we propose the use of data segmentation for phase identification and provide the theoretical foundation of the method. This significantly improves the efficiency of data usage, identification accuracy, and the robustness in the phase identification process. Second, we propose to use the CTS similarity matrix for further improving the identification accuracy by considering the relations between data participation decisions in an ensemble setting.

\section{Methodology}
\begin{figure}[t]
\centerline{\includegraphics[width=\linewidth]{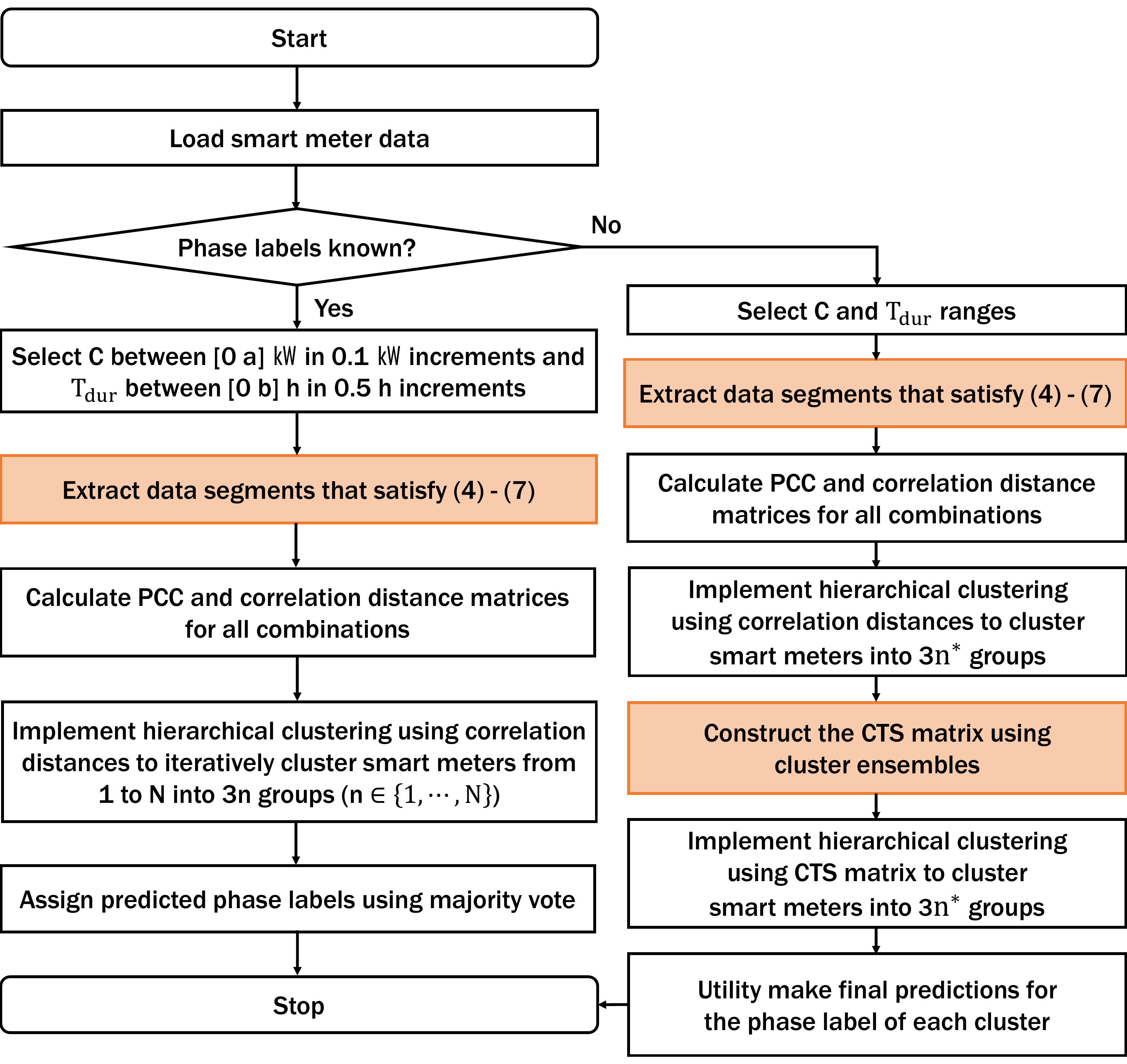}}
\vspace{-.3cm}
\caption{Flowchart of the proposed phase identification methodology.}
\label{fig1}
\vspace{-.5cm}
\end{figure}

A flowchart of the proposed data segmentation based phase identification algorithm is shown in Fig.~\ref{fig1}. Our contributions are highlighted in the three shaded boxes. The inputs of the algorithms are smart meter measurements including 15-\si{\minute} real power and voltage measurements. First, the voltage data is segmented based on customer real power consumption levels. Then, the selected data segments are used for computing Pearson correlation coefficients (PCC) and correlation distance between each pair of nodes. Based on correlation distances, the hierarchical clustering method will divide the smart meters into $3\times n$ clusters with $n$ increasing from 1 to $N$ (i.e., the number of clusters increases from 3 to 3$N$). If the phase labels are known, the majority vote mechanism presented in \cite{mitra2015voltage} will be used to assign phase labels to each cluster. If the phase labels are unknown, the CTS matrix constructed by the clustering ensemble method is proposed to determine the final clusters for utility engineers to label the phase for each cluster.

\subsection{Voltage Correlation Deterioration Phenomenon}
\begin{figure}[t]
\centerline{\includegraphics[width=\linewidth]{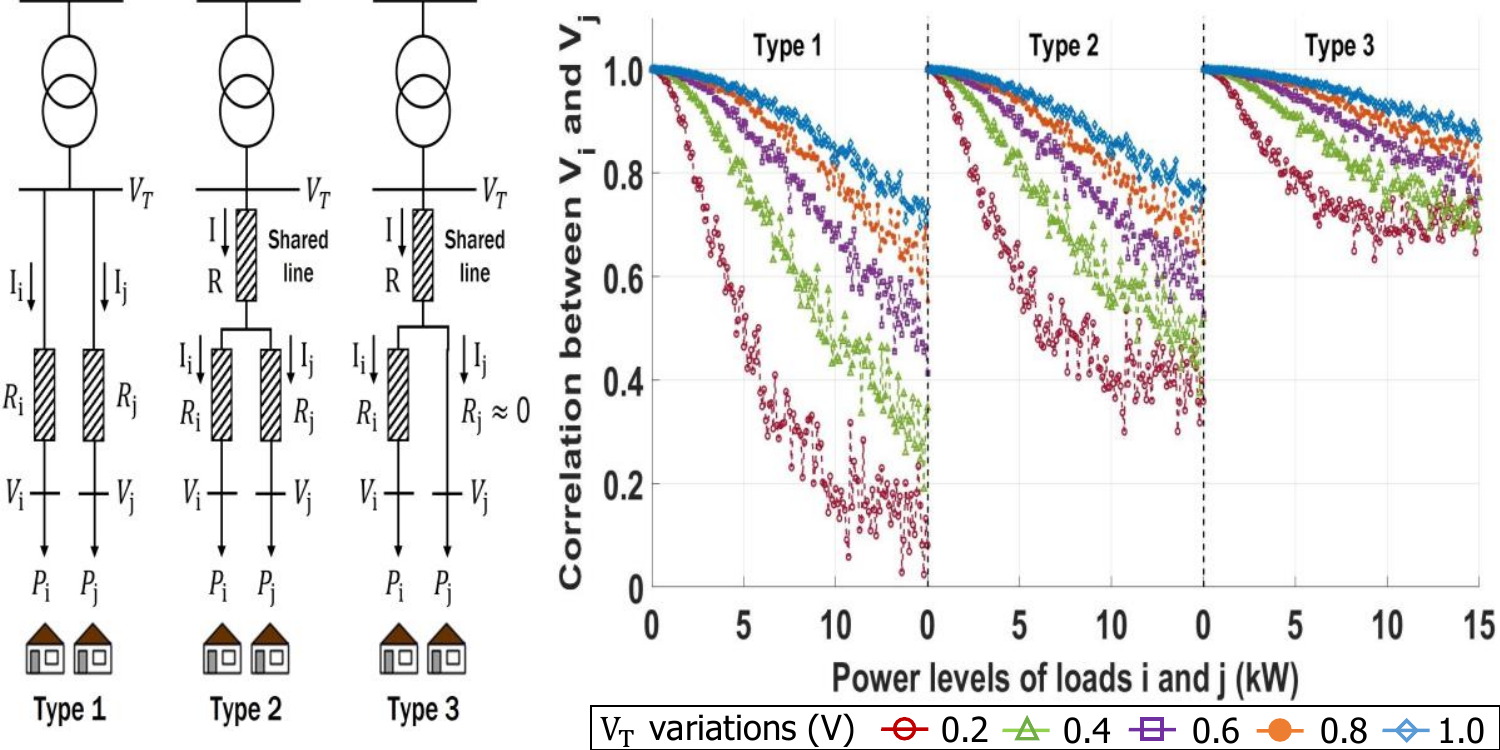}}
\vspace{-.1cm}
\subfloat[\label{1a}]{\hspace{.34\linewidth}}
\subfloat[\label{1b}]{\hspace{.66\linewidth}}
\caption{(a) Three typical types of transformer secondary circuit connections, (b) Pearson's Correlation Coefficients between the voltage of the $i^\mathrm{th}$ and $j^\mathrm{th}$ meters for types 1, 2 and 3 connections.}
\label{fig2}
\vspace{-.5cm}
\end{figure}

As shown in Fig.~\ref{fig2}(a), in relation to each other, two 1-phase loads connected to the same 1-phase transformer have three basic connection types: in-parallel, partially-parallel, and in-series. Denote $V_\mathrm{T}$, $I$, and $R$ as the distribution transformer voltage, shared line current and resistance, respectively. Denote $V_\mathrm{i}$ and $V_\mathrm{j}$, $I_\mathrm{i}$ and  $I_\mathrm{j}$ as the system voltage and current of loads $i$ and $j$, respectively. Denote $R_\mathrm{i}$ and $R_\mathrm{j}$ as the resistance of the transformer secondary connection to loads $i$ and $j$. 

Then, we have

\vspace{-.5cm}
\begin{IEEEeqnarray}{l}
V_\mathrm{i} = V_\mathrm{T} - IR - I_\mathrm{i}R_\mathrm{i} \label{eq1} \\ 
V_\mathrm{j} = V_\mathrm{T} - IR - I_\mathrm{j}R_\mathrm{j} \label{eq2} \\
I = I_\mathrm{i} + I_\mathrm{j} \label{eq3}
\end{IEEEeqnarray}
Note that for type 1, $R=0$; for type 3, $R_\mathrm{j}=0$. 

From \eqref{eq1}--\eqref{eq3}, we have the following insights. In normal operation, voltage is maintained close to its nominal values. Thus, $I_\mathrm{i}$ and  $I_\mathrm{j}$ are mainly determined by the power consumption levels of the $i^\mathrm{th}$ and $j^\mathrm{th}$ loads, $P_\mathrm{i}$ and $P_\mathrm{j}$. Therefore, if $P_\mathrm{i}$ and $P_\mathrm{j}$ are both low, then $I_\mathrm{i}$ and  $I_\mathrm{j}$ are low, causing a very small secondary voltage drop. Then, $V_\mathrm{i}$ and $V_\mathrm{j}$ will be very close to $V_\mathrm{T}$, making $V_\mathrm{i}$ and $V_\mathrm{j}$ highly correlated. However, when $P_\mathrm{i}$ and $P_\mathrm{j}$ are increasing, $I_\mathrm{i}$ and $I_\mathrm{j}$ will increase, leading to high secondary voltage drops on the secondary circuits. This will weaken the correlation between $V_\mathrm{i}$ and $V_\mathrm{j}$ significantly because the voltage variations are mainly determined by the values of $R$, $R_\mathrm{i}$, $R_\mathrm{j}$ and $|P_\mathrm{i}-P_\mathrm{j}|$.

To illustrate the phenomenon of voltage correlation deterioration with respect to the increase of local load consumption, Monte Carlo simulations are conducted to calculate PCCs between $V_\mathrm{i}$ and $V_\mathrm{j}$ for the three basic load connection types when the power levels of loads $i$ and $j$ vary randomly between [0 15] \si{\kW}; $V_\mathrm{T}$ varies randomly from 120 V within five bands; and $R$, $R_\mathrm{i}$, and $R_\mathrm{j}$ are set as 0.01 \si{\ohm}, 0.05 \si{\ohm}, and 0.05 \si{\ohm}, respectively. 

As shown in Fig.~\ref{fig2}(b), the voltage correlation deterioration for the type 1 connection is the most prominent. When the output levels of both loads increase above 5 kW while the variation of $V_\mathrm{T}$ is small (e.g., 0.2 V against 120 V), the correlation can drop below 0.4. The voltage correlation deterioration can exacerbate when the two loads are served by different distribution transformers on the same phase due to the increase of the line impedance. Therefore, excluding those data sets that cause correlation deterioration is crucial for improving the accuracy of voltage correlation based phase identification methods.

\subsection{Data Segmentation Algorithm}
In \cite{pezeshki2012consumer, olivier2018phase}, the authors have proven that using voltage correlation for clustering is an effective method for customer phase identification. Based on the insight gained from the circuit analysis in Section~\rom{2}.A, a data segmentation algorithm is developed for extracting data segments that satisfy the low power and minimum data length conditions to obtain data segments with strong correlation between two time series $V_\mathrm{i}$ and $V_\mathrm{j}$ if they are on the same phase. This process can be formulated as

\vspace{-.5cm}
\begin{IEEEeqnarray}{lCr}
0 \leq P_\mathrm{i,t} \leq C, \quad T_\mathrm{dur} \leq m_\mathrm{i,k}\Delta T  \label{eq4} \\ 
0 \leq P_\mathrm{j,t} \leq C, \quad T_\mathrm{dur} \leq m_\mathrm{j,k}\Delta T  \label{eq5} \\ 
PCC(V^\mathrm{M}_\mathrm{i}, V^\mathrm{M}_\mathrm{j}) = \nonumber \\ \hfill \frac{\sum_{k=1}^{K}(V^\mathrm{m_\mathrm{k}}_\mathrm{i}-\overline{V}^\mathrm{M}_\mathrm{i})(V^\mathrm{m_\mathrm{k}}_\mathrm{j}-\overline{V}^\mathrm{M}_\mathrm{j})}{\sqrt{\sum_{k=1}^{K}(V^\mathrm{m_\mathrm{k}}_\mathrm{i}-\overline{V}^\mathrm{M}_\mathrm{i})^\mathrm{2}}{\sqrt{\sum_{k=1}^{K}(V^\mathrm{m_\mathrm{k}}_\mathrm{j}-\overline{V}^\mathrm{M}_\mathrm{j})^\mathrm{2}}}} \label{eq6} \\ 
D(V^\mathrm{M}_\mathrm{i}, V^\mathrm{M}_\mathrm{j}) = 1 - \left| PCC(V^\mathrm{M}_\mathrm{i}, V^\mathrm{M}_\mathrm{j}) \right| \label{eq7}
\end{IEEEeqnarray}
where $i$ and $j$ are the indices of voltage measurement ($i,j \in \{1,\cdots,N_\mathrm{M}\}$), $t$ is the time steps ($t \in \{1,\cdots,T\}$), $\Delta T$ is the sampling interval, $C$ is the power threshold, $T_\mathrm{dur}$ is the minimum duration, $\overline{V}^\mathrm{M}_\mathrm{i}$ and $\overline{V}^\mathrm{M}_\mathrm{j}$ are the mean values of $V^\mathrm{M}_\mathrm{i}$ and $V^\mathrm{M}_\mathrm{j}$, $k$ is the index of the data segment ($k \in \{1,\cdots,K\}$), $m_\mathrm{k}$ is the number of data points in the $k^\mathrm{th}$ data segments, $M = \{m_\mathrm{1},\cdots,m_\mathrm{K}\}$ is the set of the data segments. Note that \eqref{eq4} and \eqref{eq5} are the data segmentation criteria (i.e., power threshold and minimum duration); \eqref{eq6} computes the PCC matrix; \eqref{eq7} computes the correlation distance.

\begin{figure}[t]
\centerline{\includegraphics[width=\linewidth]{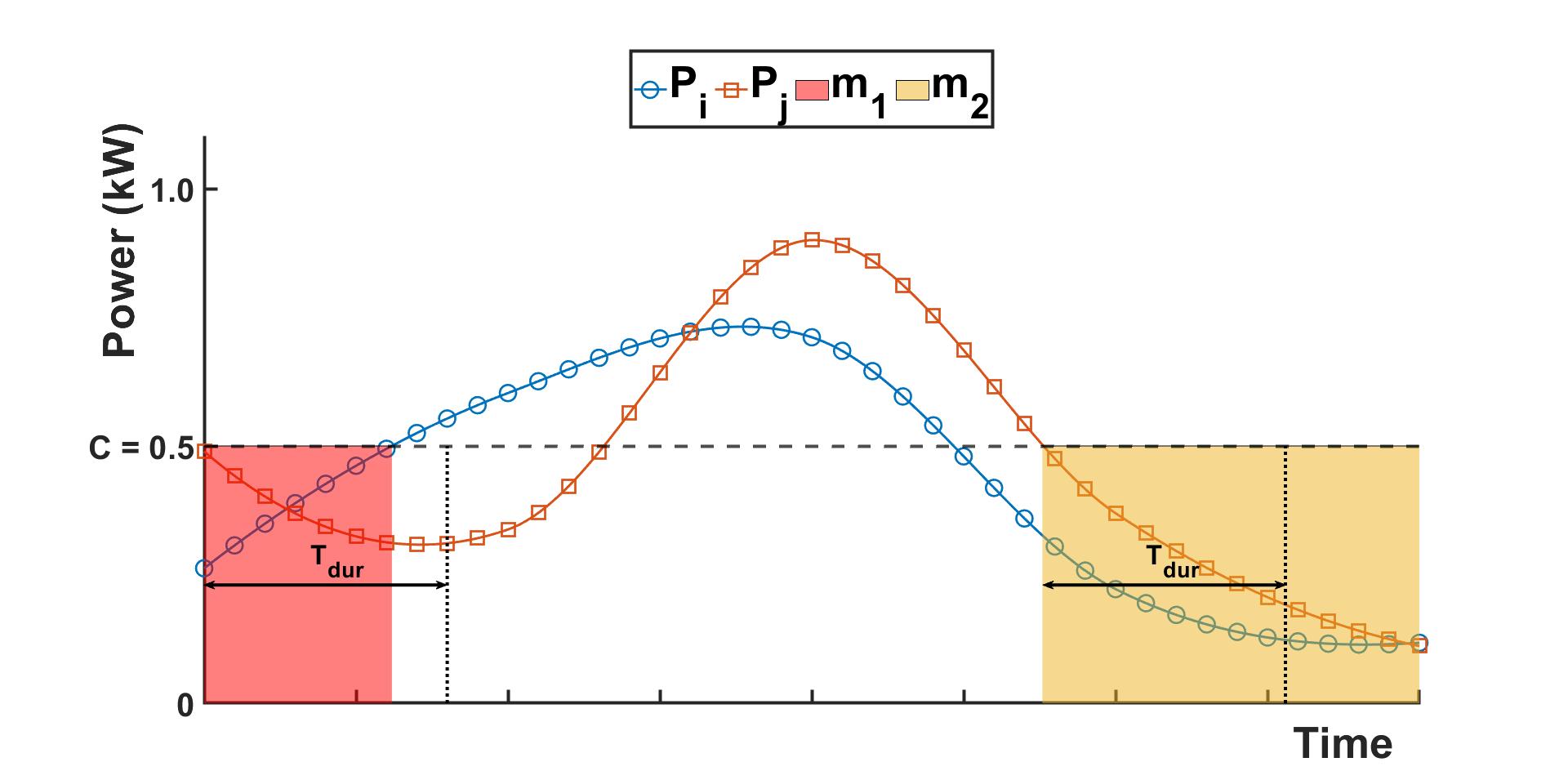}}
\vspace{-.3cm}
\caption{Data segments selection between the real power of the $i^\mathrm{th}$ and $j^\mathrm{th}$ meters.}
\label{fig3}
\vspace{-.5cm}
\end{figure}

As shown in Fig.~\ref{fig3}, only the line segments below the power consumption $C$ and the number of data points in the line segments that exceed $T_\mathrm{dur}$ are selected. The minimum duration constraint is added because when there are not enough data points in the selected line segments, the calculation of correlation between two time-series data sets is no longer meaningful. If no data segment satisfies the low-power condition, all data is used to compute the PCC. The selection of the best $C$ and $T_\mathrm{dur}$ values will be further discussed in Section~\rom{3}.

\subsection{Clustering Methods}
The hierarchical clustering method introduced in \cite{johnson1967hierarchical} is used to partition loads into $3\times n$ clusters based on the correlation distance calculated from the PCC matrix. Note that the loads cannot simply be grouped into three clusters, representing the $a$, $b$, and $c$ phases, respectively. Our circuit analysis in Section~\rom{2}.A has shown that for two loads on parallel lateral circuits, even though they are on the same phase, the voltage correlation between the two loads can be low due to high $R_{\mathrm{i}}$ and $R_{\mathrm{j}}$. Therefore, the optimal number of clusters vary from circuit to circuit. In the results section, we will discuss the selection of the number of clusters. Once the final number of clusters is determined, a phase label needs to be assigned to each cluster so that all the loads inside the cluster will be on the same phase. If the utility has phase labels for each load, a majority vote method can be used to determine the phase label for each cluster. Assigned labels are evaluated for prediction accuracy according to the percentage of customers assigned correct labels.

If phase labels are not available to the analyst, the goal of the phase identification problem changes to: for a given number of clusters, grouping customers most likely on the same phase together in a cluster. This allows utility engineers to label the phase for each cluster instead of for each customer, which, consequently, reduces the number of field inspections from hundreds/thousands to tens of sites.

In \cite{blakely2020phase}, Blakely \textit{et al}. proposed a method to obtain the final clusters using co-association matrix-based ensemble clustering. However, 'k-mean' or 'discretise' based spectral clustering is sensitive to initialization. The co-association matrix can be less accurate because, when estimating similarity, it only considers customers assigned to the same cluster.  Therefore, in this paper, we adapted the CTS matrix ensemble clustering model introduced in \cite{iam2011link} to determine members in each cluster. The CTS matrix estimates the similarity between customers by considering not only pairwise similarity, but also the relations between data participation decisions for each scenario in an ensemble. Details regarding the CTS method can be found in \cite{iam2011link}. 

\begin{figure}[t]
\centerline{\includegraphics[width=\linewidth]{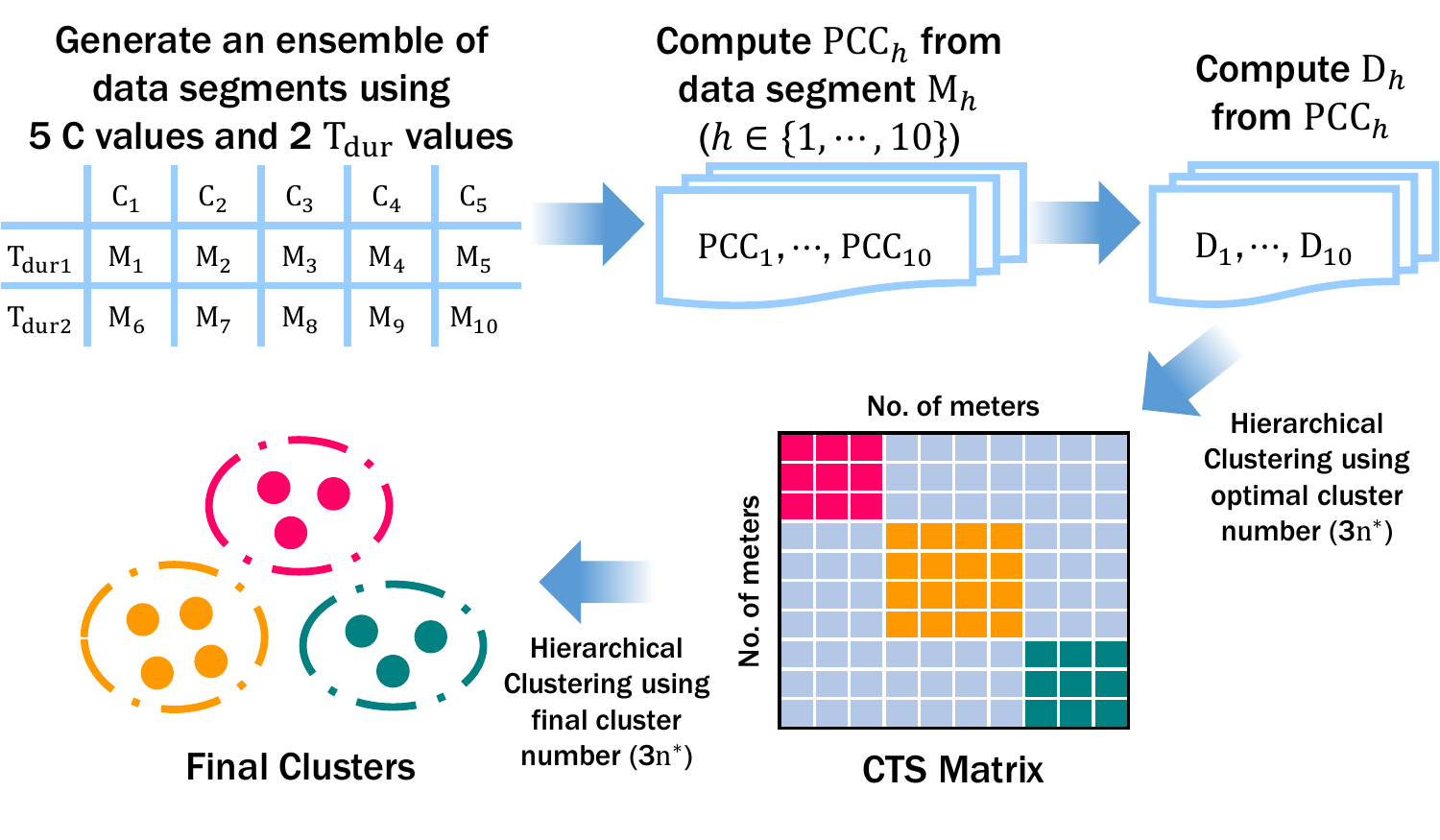}}
\vspace{-.3cm}
\caption{Process of the CTS matrix ensemble clustering.}
\label{fig4}
\vspace{-.5cm}
\end{figure}

As illustrated in Fig. \ref{fig4}, an ensemble of the PCC matrix can be constructed by selecting different values for $C$ and $T_\mathrm{dur}$. For example, select 5 $C$ values and 2 $T_\mathrm{dur}$ values can yield 10 segmented time-series data sets (i.e., $M_{\mathrm{1}}, ..., M_{\mathrm{10}}$). Then, calculate PCC and correlation distance matrices for each of the 10 segmented data sets. Note that each of the 10 correlation distance matrix can be partitioned into a targeted cluster number, 3$n^\mathrm{*}$, (in the paper, 3$n^\mathrm{*}$ is set to be 36 clusters) using hierarchical clustering. Next, the CTS matrix is constructed for estimating the similarity of the 10 sets of smart meters assigned to the 36 clusters. Finally, hierarchical clustering is used on the CTS matrix to obtain the final members in the 36 clusters.

\section{Simulation Results}
Two types of data sets are used for developing and testing the proposed algorithm. The first data set is synthetic data and uses a modified IEEE 123-bus system to generate one year of 15-\si{\minute} data for 1,100 loads. The feeder load disaggregation algorithm presented in \cite{wang2020data} has been used to allocate 1-\si{\minute} resolution residential load profiles from Pecan Street \cite{PECAN} to every load node on a test feeder. Each load node in 123 bus system is assigned a minimum of 4, a maximum of 26, and an average of 11 loads.  

The second data set includes one year of 15-\si{\minute} data for 1,448 smart meters collected from three distribution feeders by a local utility in North Carolina. The data sets contain 8.57\% missing data spread throughout customers and customers with more than 80\% missing data are removed from the data set. To verify the performance of the proposed algorithm by data length, one year data from Feeder 1 and three months data from Feeder 2 and 3 are used. The voltages are normalized based on their service voltage levels. The phase of each 1-phase distribution transformer is provided by the utility, subsequently, the phase label of each meter is the same as that of the transformer it connected to.

\subsection{Case 1: With Known Customer Phase Labels}
\begin{figure}[t]
\centerline{\includegraphics[width=\linewidth]{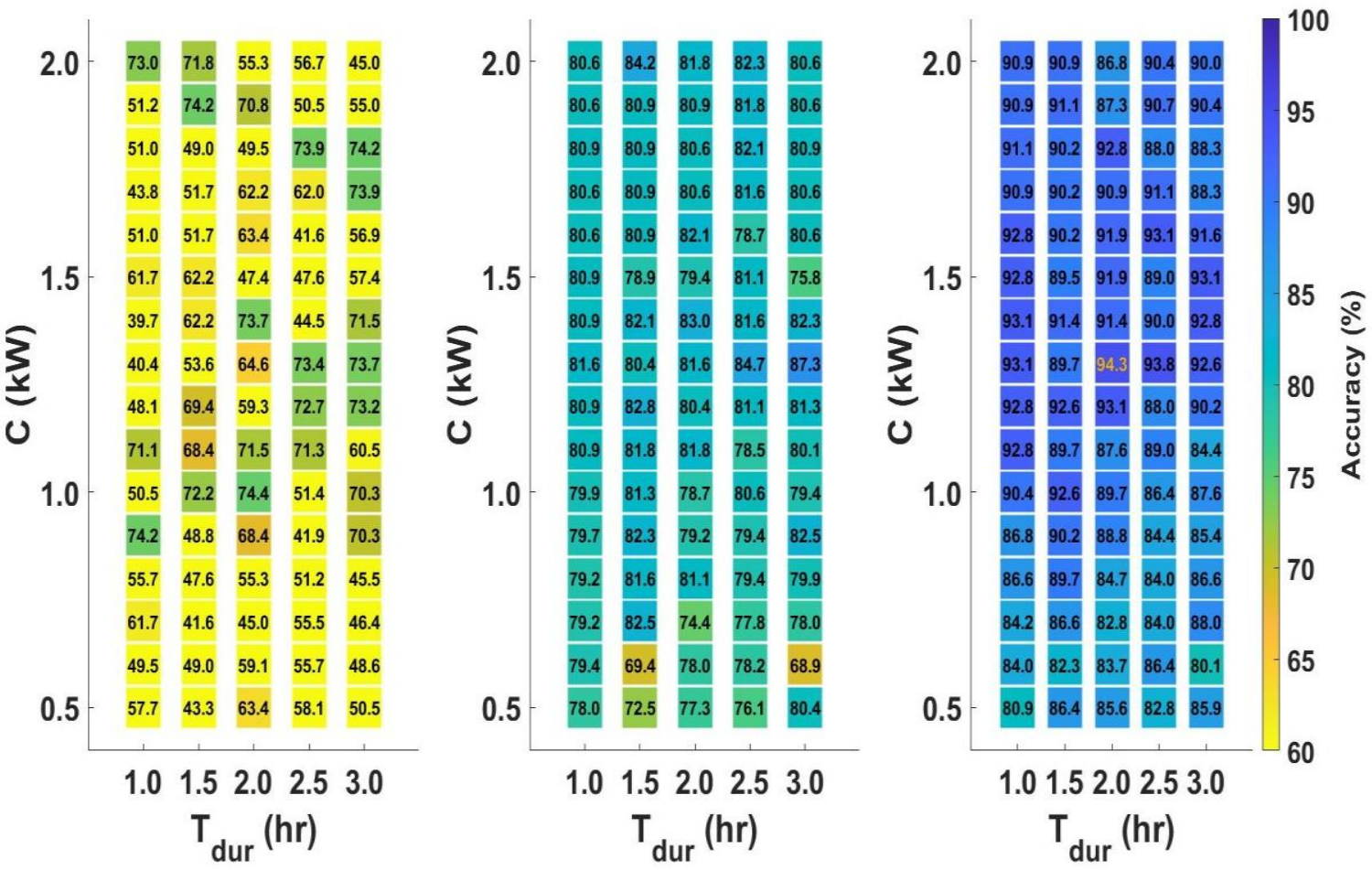}}
\vspace{-.3cm}
\subfloat[3 clusters \label{5a}]{\hspace{.35\linewidth}}
\subfloat[18 clusters \label{5b}]{\hspace{.32\linewidth}}
\subfloat[36 clusters \label{5c}]{\hspace{.33\linewidth}}
\caption{Phase identification accuracy for three different numbers of clusters with varying parameters in Feeder~2. Maximum accuracy obtained when $C=1.3$ \si{\kW}, $T_\mathrm{dur}=2.0$ \si{\hour}, and $3n^*=36$.}
\label{fig5}
\vspace{-.5cm}
\end{figure}

To select the optimal parameters for each feeder, we conduct Monte Carlo simulations for the range of parameters shown in Table \ref{tab1}. As shown in Fig.~\ref{fig5}, if $C$ and $T_\mathrm{dur}$ are small, there are very few data segments satisfying the low-power and minimum data length conditions. This renders the correlation calculation unreliable. As shown in Table \ref{tab1}, the maximum accuracy occurs when $C\geq 0.7$ \si{\kW} and $T_\mathrm{dur}\geq 1$ \si{\hour}. Therefore, parameter ranges $C$ between [0 0.4] \si{\kW} and $T_\mathrm{dur}$ between [0 0.5] \si{\hour} are removed from the subsequent simulations. 

The results also indicate that the optimal number of clusters varies from circuit to circuit. In our simulation, Feeder 1, a short feeder with 131 customers, requires only 9 clusters; Feeders 2 (418 customers) and 3 (899 customers), feeders with long laterals, require 36 clusters. This suggests that for circuits with long laterals, clustering customers to more groups is necessary. As discussed in type I circuit analysis in Section~\rom{2}.A, voltage correlation can be weakened for a pair of meters that are on the same phase but supplied by different lateral circuits because of the large $R_{\mathrm{i}}$ and $R_{\mathrm{j}}$. 

\begin{table}[t]
\renewcommand{\arraystretch}{1.3}
\caption{Parameter Selection of Synthetic and Real Feeders}
\vspace{-.2cm}
\label{tab1}
\centering
\begin{tabularx}{\columnwidth}{lccccc}
\hline
\multirow{2}{4em}{\parbox{1\linewidth}{\vspace{0.2cm} Parameters}} &\multirow{2}{4em}{\parbox{1\linewidth}{\vspace{0.2cm} Parameter ranges}} &\multicolumn{4}{c}{Optimal values} \\ \cmidrule{3-6}
&   &Synthetic   &Feeder 1   &Feeder 2   &Feeder 3 \\ \hline \hline
\textbf{$C$ \si{[\kW]}}       &[0 2.0]  &1.0  &0.7, 1.0  &1.3  &1.2  \\ 
\textbf{$T_\mathrm{dur}$ \si{[\hour]}}  &[0 3.0]  &0.5  &[1.0 3.0]  &2.0  &2.0 \\
\textbf{$3\times n$}            &[3 36]   &36  &[9 36]  &36  &36       \\ \hline
\end{tabularx}
\vspace{-.3cm}
\end{table}

\begin{figure}[t]
\centerline{\includegraphics[width=\linewidth]{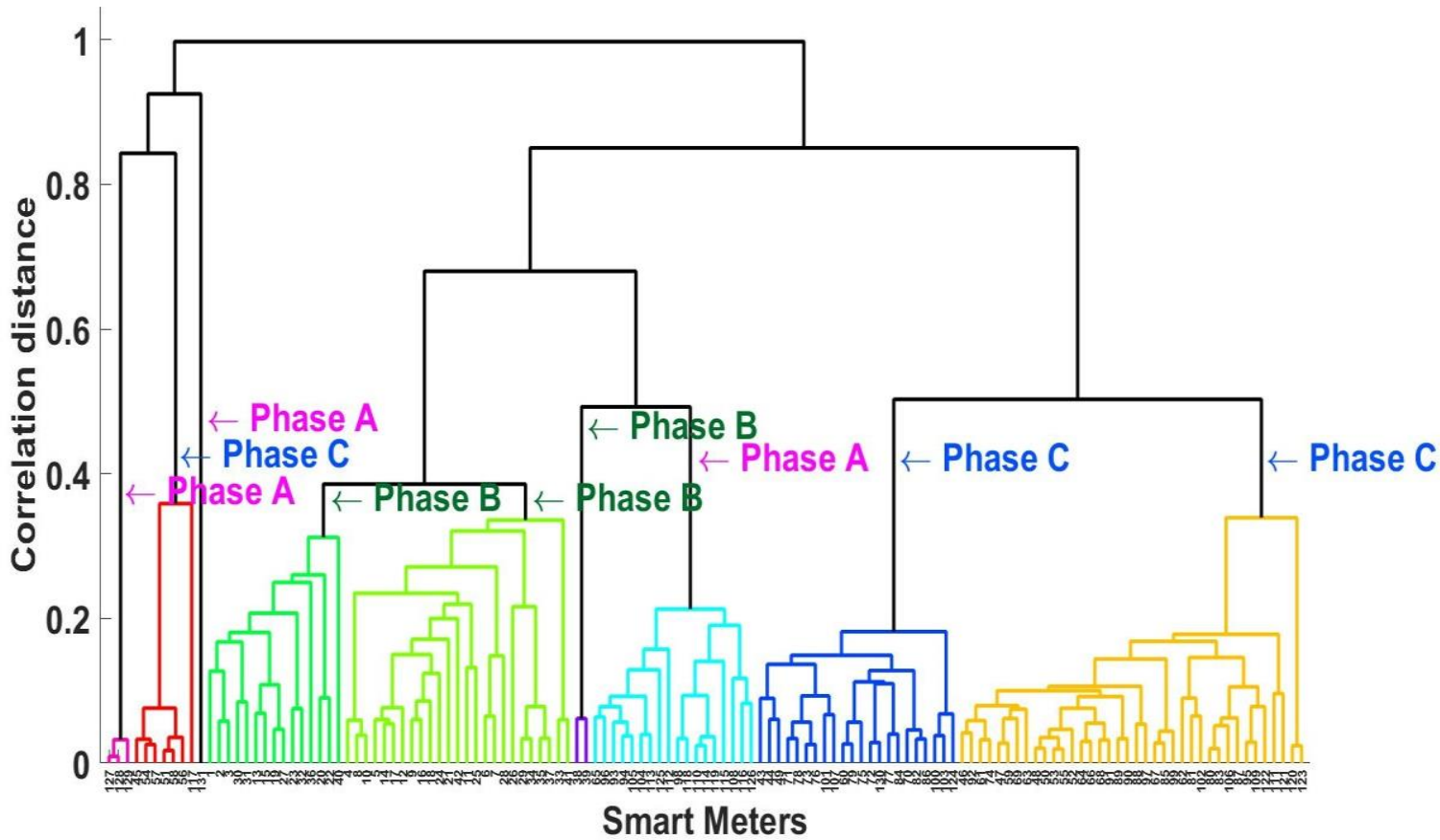}}
\vspace{-.3cm}
\caption{Hierarchical clustering result of real feeder~1}
\label{fig6}
\vspace{-.1cm}
\end{figure}

Table \ref{tab2} reports the phase identification results. The proposed method predicts 100\% of the original utility phase labels in synthetic and Feeder 1 data sets. Figure~\ref{fig6} shows the hierarchical clustering results for real feeder 1. For Feeders 2 and 3, we obtained 94.2\% and 85.7\% matching using the original utility phase labels. After site verification, 2.4\% and 12.5\% of "mislabeled" data proved to be correct. 

For the remaining 3.4\% and 1.9\% mislabeled customers on Feeders 2 and 3, there are two observations. First, many mislabeled customers are locate at the end of the feeder, where the voltage correlation deteriorates significantly because of the high line impedance (See analysis in Section~\rom{2}.A). Second, a few mislabeled customers have very high power consumption so that very few data segments can meet the low-power and minimum data length conditions, making the calculation of correlation unreliable.

\subsection{Case 2: Without Known Customer Phase Labels}

To generate the ensembles for the without-phase-label cases, for the synthetic feeder, we vary $C$ between [0.4 0.8] \si{\kW} with 0.1 \si{\kW} increment, select $T_\mathrm{dur}$ to be 2.5 or 3.0 \si{\hour}, set the number of clusters to be 36. For the three real feeder data sets, we vary $C$ between [1.2 1.6] \si{\kW} with 0.1 \si{\kW} increment, select $T_\mathrm{dur}$ to be 2.0 or 2.5 \si{\hour}, and let 3$n^\mathrm{*}=36$.

The proposed method is compared with the ensemble spectral clustering using the GIS topology (ESC-GIS) presented in \cite{blakely2020phase}. To run ESC-GIS, we set the window size to 4 days and the number of clusters to 6, 12, 15, and 30. As shown in Table \ref{tab3}, the proposed method outperformed ESC-GIS for all four cases. This is because ESC-GIS generates cluster ensembles by combining different numbers of clusters, so the results can be inaccurate unless the optimal number of clusters for each data set is accounted for.

\begin{table}[t!]
\caption{Case~1: Phase Identification Results}
\vspace{-.2cm}
\label{tab2}
\centering
\begin{tabularx}{\columnwidth}{p{2cm}cccc}
\hline
 & Synthetic          & Feeder~1          & Feeder~2          & Feeder~3   \\ \hline \hline
 \hspace{-.2cm} Phase A\\
 Recorded phase             & 436   & 22   & 163   & 344  \\
 Predicted phase            & 436   & 22   & 160   & 339  \\
\hline 
\hspace{-.2cm} Phase B\\
 Recorded phase             & 293   & 42   & 96   & 249  \\
 Predicted phase            & 293   & 42   & 90   & 255  \\  
\hline 
 \hspace{-.2cm} Phase C\\
 Recorded phase             & 371   & 67   & 159   & 306  \\
 Predicted phase            & 371   & 67   & 168   & 305  \\  
\hline
\hspace{-.2cm} Total~($N_\mathrm{T}$)   & 1,100   & 131   & 418            & 899  \\
 Corrected~($N_\mathrm{c}$)             & N/A     & 0     & 10 (2.4\%)     & 112 (12.5\%) \\
 Validated~($N_\mathrm{v}$)             & 1,100   & 131   & 394 (94.3\%)   & 770 (85.7\%) \\
   \multirow{2}{4em}{\parbox{1\linewidth}{\vspace{.01cm} \textbf{Accuracy} (($N_\mathrm{c}$+$N_\mathrm{v}$)/$N_\mathrm{T}$)}}   &\multirow{2}{1em}{\textbf{100\%}}   &\multirow{2}{1.5em}{\textbf{100\%}}   &\multirow{2}{1.5em}{\textbf{96.6\%}}  &\multirow{2}{1.5em}{\textbf{98.1\%}} \\ \\
\hline
\end{tabularx}
\vspace{-.4cm}
\end{table}

\begin{table}[t!]
\caption{Case~2: Phase Identification Results and Performance Comparison}
\vspace{-.2cm}
\label{tab3}
\centering
\begin{tabularx}{\columnwidth}{p{2cm}cccc}
\hline
 & Synthetic          & Feeder~1          & Feeder~2          & Feeder~3   \\ \hline \hline
 \hspace{-.2cm} Phase A\\
 Recorded phase             & 436   & 22   & 163   & 344  \\
 Predicted phase            & 436   & 22   & 167   & 327  \\ 
\hline 
\hspace{-.2cm} Phase B\\
 Recorded phase             & 293   & 42   & 96   & 249  \\
 Predicted phase            & 295   & 42   & 83   & 252  \\  
\hline 
 \hspace{-.2cm} Phase C\\
 Recorded phase             & 371   & 67   & 159   & 306  \\
 Predicted phase            & 369   & 67   & 168   & 320  \\  
\hline
\hspace{-.2cm} \textbf{Proposed Method}\\
Corrected ($N_\mathrm{c}$)    & N/A     & 0     & 10 (2.4\%)    & 112 (12.5\%)  \\
 Validated ($N_\mathrm{v}$)   & 1,098   & 131   & 393 (94.0\%)  & 784 (87.2\%)  \\
 Total ($N_\mathrm{T}$)       & 1,100   & 131   & 418           & 899  \\
 \multirow{2}{4em}{\parbox{1\linewidth}{\vspace{.01cm} \textbf{Accuracy} (($N_\mathrm{c}$+$N_\mathrm{v}$)/$N_\mathrm{T}$)}}   &\multirow{2}{1em}{\textbf{99.8\%}}   &\multirow{2}{1.5em}{\textbf{100\%}}   &\multirow{2}{1.5em}{\textbf{96.4\%}}  &\multirow{2}{1.5em}{\textbf{99.7\%}} \\ \\
\hline
\hspace{-.2cm} \textbf{ESC-GIS \cite{blakely2020phase}}\\
 Corrected~($N_\mathrm{c}$)   & N/A     & 0     & 10 (2.4\%)     & 112 (12.5\%) \\
 Validated~($N_\mathrm{v}$)   & 1,068   & 130   & 385 (92.1\%)   & 780 (86.7\%) \\
 Total~($N_\mathrm{T}$)       & 1,100   & 131   & 418            & 899  \\
 \multirow{2}{4em}{\parbox{1\linewidth}{\vspace{.01cm} \textbf{Accuracy} (($N_\mathrm{c}$+$N_\mathrm{v}$)/$N_\mathrm{T}$)}}   &\multirow{2}{1em}{\textbf{97.0\%}}   &\multirow{2}{1.5em}{\textbf{99.2\%}}   &\multirow{2}{1.5em}{\textbf{94.5\%}}  &\multirow{2}{1.5em}{\textbf{99.2\%}} \\ \\
\hline
\end{tabularx}
\vspace{-.5cm}
\end{table}

\section{Conclusion}
In this paper, a novel data-segmentation method for customer phase identification is presented. First, we apply circuit analysis to demonstrate the existence and cause of voltage correlation deterioration phenomena for three typical connection types of a pair of 1-phase customers on the same phase. This inspired us to use low-power threshold and minimum data duration for extracting highly correlated data segments for computing voltage correlations instead of using all available voltage measurements. Simulation results on one synthetic feeder and three actual feeders show that the proposed algorithm can significantly improve the accuracy of customer phase identification. When customer phase labels are unknown, we propose to use CTS-based similarity matrix for hierarchical clustering to account for both the pairwise similarity and the hidden relations between data partitions for each scenario in an ensemble to further improve the clustering accuracy. Our future work will be to test the algorithm on more utility feeders and improve the algorithm performance on customers at the end of the feeder or with large power consumption.

\ifCLASSOPTIONcaptionsoff
\newpage
\fi 
\bibliographystyle{IEEEtran}
\bibliography{IEEEabrv,MyRefs}

\end{document}